\documentclass[sigconf, nonacm]{acmart}
\pdfmapfile{pdftex_ndl14.map}
\usepackage{graphicx}
\usepackage{balance}
\usepackage{bbm}
\usepackage{url}
\usepackage{amsmath}
\usepackage{subcaption}
\usepackage{tcolorbox}
\usepackage{cleveref}
\usepackage{multirow}

\usepackage{comment}
\usepackage{enumitem}
\usepackage{xspace}

\newcommand\vldbpages{2735 - 2738}
\newcommand\vldbpagestyle{empty}
\newcommand\vldbvolume{14}
\newcommand\vldbissue{12}
\newcommand\vldbyear{2021}
\newcommand\vldbauthors{\authors}
\newcommand\vldbtitle{\shorttitle} 
\newcommand\vldbavailabilityurl{}
\newcommand\vldbdoi{10.14778/3476311.3476332}

\newcommand{\at}[1]{\protect\ensuremath{\mathsf{#1}}\xspace}

\newcommand{\systemx}{\at{Panda}}
\newcommand{\stitle}[1]{\vspace{1ex}\noindent{\bf #1}}

\newcommand\revise[1]{\textcolor{black}{#1}}

\begin{document}
\title{Demonstration of Panda: \\ A Weakly Supervised Entity Matching System}

\author{
Renzhi Wu$^{\dagger}$,
Prem Sakala$^{\dagger}$,
Peng Li$^{\dagger}$,
Xu Chu$^{\dagger}$,
Yeye He$^{\mathsection}$,
}
\affiliation{
\institution{$^\dagger$Georgia Institute of Technology, $^\mathsection$Microsoft Research}
\institution{$^\dagger$\{renzhiwu@, premsakala@, pengli@, xu.chu@cc.\}gatech.edu, $^\mathsection$yeyehe@microsoft.com}
}

\renewcommand{\shortauthors}{R. Wu, P. Sakala, P. Li, X. Chu, Y. He}
\renewcommand{\authors}{
Renzhi Wu,
Prem Sakala,
Peng Li,
Xu Chu,
Yeye He}







\begin{abstract}
Entity matching (EM) refers to the problem of identifying tuple pairs in one or more relations that refer to the same real world entities. Supervised machine learning (ML) approaches, and deep learning based approaches in particular, typically achieve state-of-the-art matching results. However, these approaches require many labeled examples, in the form of matching and non-matching pairs, which are expensive and time-consuming to label. 



In this paper, we introduce \systemx, a weakly supervised system specifically designed for EM. \systemx uses the same \textit{labeling function} abstraction as Snorkel, where labeling functions (LF) are user-provided programs that can generate large amounts of (somewhat noisy) labels quickly and cheaply, which can then be combined via a labeling model to generate accurate final predictions. To support users developing LFs for EM, \systemx provides an integrated development environment (IDE) that lives in a modern browser architecture. \systemx's IDE facilitates the development, debugging, and life-cycle management of LFs in the context of EM tasks, similar to how IDEs such as Visual Studio or Eclipse excel in general-purpose programming. \systemx's IDE includes many novel features purpose-built for EM, such as smart data sampling, a builtin library of EM utility functions, automatically generated LFs, visual debugging of LFs, and finally, an EM-specific labeling model. We show in this demo that \systemx IDE can greatly accelerate the development of high-quality EM solutions using weak supervision.





\end{abstract}

\maketitle
\renewcommand{\vldbtitle}{
Demonstration of Panda: A Weakly Supervised Entity Matching System}

\vspace{-1mm}
\pagestyle{\vldbpagestyle}
\begingroup\small\noindent\raggedright\textbf{PVLDB Reference Format:}\\
\vldbauthors. \vldbtitle. PVLDB, \vldbvolume(\vldbissue): \vldbpages, \vldbyear.\\
\href{https://doi.org/\vldbdoi}{doi:\vldbdoi}
\endgroup
\begingroup
\renewcommand\thefootnote{}\footnote{\noindent
This work is licensed under the Creative Commons BY-NC-ND 4.0 International License. Visit \url{https://creativecommons.org/licenses/by-nc-nd/4.0/} to view a copy of this license. For any use beyond those covered by this license, obtain permission by emailing \href{mailto:info@vldb.org}{info@vldb.org}. Copyright is held by the owner/author(s). Publication rights licensed to the VLDB Endowment. \\
\raggedright Proceedings of the VLDB Endowment, Vol. \vldbvolume, No. \vldbissue\ %
ISSN 2150-8097. \\
\href{https://doi.org/\vldbdoi}{doi:\vldbdoi} \\
}\addtocounter{footnote}{-1}\endgroup

\ifdefempty{\vldbavailabilityurl}{}{
\vspace{.3cm}
\begingroup\small\noindent\raggedright\textbf{PVLDB Artifact Availability:}\\
The source code, data, and/or other artifacts have been made available at \url{\vldbavailabilityurl}.
\endgroup
}
\vspace{-1mm}
\section{Introduction}

Entity matching (EM) refers to the problem of identifying tuples in one or more tables that refer to the same real world entities.
For example, an e-commerce website would want to identify identical products from different suppliers for a unified catalog.
EM has been extensively studied in many research communities, including databases, statistics, NLP, and data mining. The standard approach to EM uses similarity scores between two tuples as feature vectors, and then formulates the problem of match/non-match as a binary classification problem given the feature vectors ~\cite{DBLP:journals/tkde/ElmagarmidIV07}.
Supervised Machine Learning (ML) approaches, particularly deep learning approaches (e.g., DeepER~\cite{deeper} and DeepMatcher~\cite{anhaisigmod2018}), often achieve state-of-the-art results for EM~\cite{deeper,anhaisigmod2018,kopcke2010evaluation,dong2018data}.
However, they require large amounts of labeled training data that are expensive and time-consuming to obtain -- for example, it has been reported~\cite{dong2018data} that achieving F-measures of $\sim$99\% with random forests can
require up to 1.5M labels even on relatively clean datasets.

\vspace{-1mm}
\stitle{Weak supervision and the LF abstraction.} Lack of training data is a common problem in applied ML. Many ML researchers and practitioners have increasingly resorted to weak supervision methods, in which large amounts of cheaply generated, but often noisy, labeled examples are generated in lieu of hand-labeled examples. 


The data programming paradigm~\cite{ratner2016data}, as implemented in the Snorkel system~\cite{ratner2017snorkel}, 
proposes to generate weakly supervised signals using labeling functions (LFs). LFs are user-provided functions (e.g., in Python) that take each example as input and produces a possibly noisy label (positive/negative/abstain). 
A labeling model then combines noisy labels from all LFs,  considers their accuracy and possible correlations, to produce a final probabilistic label. Data programming has been successfully adopted across many industries and application domains.



\vspace{-1mm}
\stitle{Weak supervision for EM.} We built a new weak supervision system, \systemx, that is specifically designed for EM. \systemx adopts the data programming paradigm, and develops an IDE that allows users to easily develop LFs in order to build weakly-supervised solutions for EM tasks.

\begin{figure}[htb!]
	\centering
	\vspace{-2mm}	\includegraphics[width=0.9\linewidth]{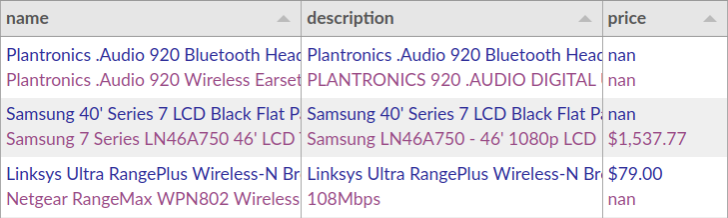}
	\vspace{-4mm}
	\caption{\small{Tuple pairs in the abt-buy dataset~\cite{erhardws}. Blue denotes tuples from the abt table and purple denotes tuples from the buy table.}} 
	\label{fig:abt_buy_tuple_pairs} 
	\vspace{-6mm}
\end{figure}

\begin{figure}[htb!]
	\centering
\vspace{-4mm}	
\includegraphics[width=\linewidth]{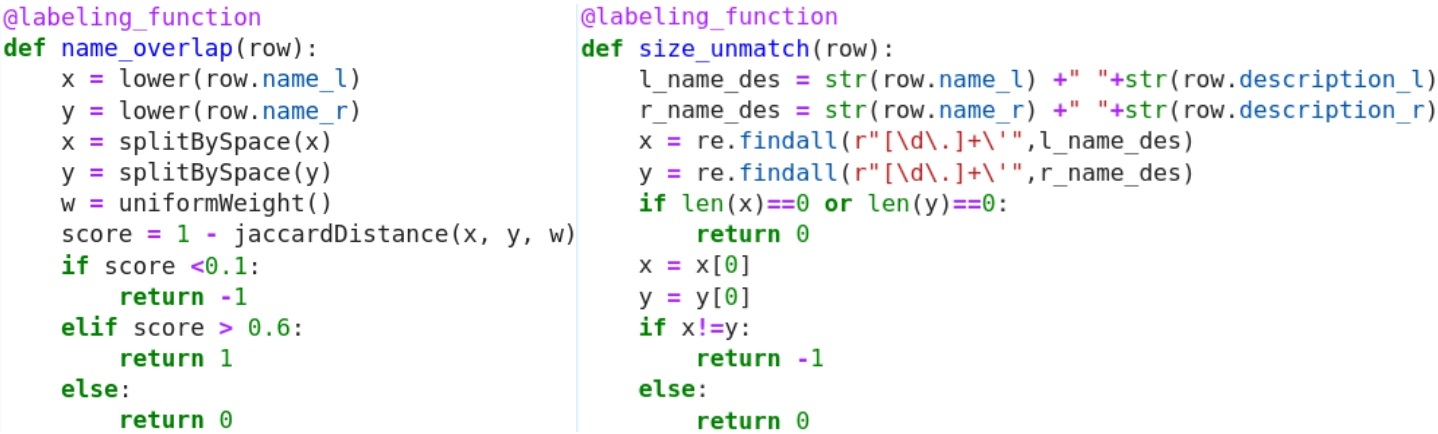}
	\vspace{-7mm}
	\caption{Example labeling functions for the abt-buy dataset} 
	\label{fig:abt_buy_lfs} 
	\vspace{-3mm}
\end{figure}

\begin{example}
Figure~\ref{fig:abt_buy_tuple_pairs} shows a sample of tuple pairs for the abt-buy dataset~\cite{erhardws}. Instead of manually labeling each tuple pair, users may inspect the record pairs and develop some intuition of how to label a pair as match vs. non-match. For example, one may label a tuple pair as a match if the "name" attribute of the two products are very similar (e.g. the first pair). On the other hand, a tuple pair is likely a non-match if key attributes of the two products are different (e.g. the second tuple pair, where the screen sizes are 40' vs 46').  


Users can encode these intuitions in the form of LFs.  Figure~\ref{fig:abt_buy_lfs} shows two example LFs written in Python that correspond to the example heuristics mentioned above. 
The {\at{name\_overlap}} function labels a tuple pair a match (+1) or non-match (-1) if the token overlap of their "name" attribute is high (score > 0.6) or low (score < 0.1); otherwise, it abstains (0).

As a second example of LF, the {\at{size\_unmatch}} function on the right uses a regular expression to extract the product size (e.g. 40') from the "name" and "description" attributes. It labels a tuple pair a non-match (-1) if the sizes are different, and abstains (0) otherwise. 
As one can imagine, LFs so created may not be perfectly accurate and may have varying levels of coverage. Nevertheless, as long as LFs are reasonably accurate (e.g., better than random labeling), it has been shown~\cite{ratner2016data} that a labeling model can reason about the accuracy/correlation of LFs to produce accurate final predictions. 
\end{example}

Since the quality of final labels depends critically on the user-provided LFs, \textit{\systemx aims to provide an integrated development environment (IDE) to support users in the entire life-cycle of labeling using LFs for EM tasks.} Just like a general-purpose IDE (e.g., Visual Studio or Eclipse) that would include many features to support users write and debug programs, \systemx's IDE includes many features to support users to write, debug, and combine LFs via a browser-based GUI that is both contextual and intuitive (Section 2).

\vspace{-1mm}
\stitle{Related Work.} Most existing work in the literature focuses on optimizing individual components of the EM pipeline (e.g., blocking and matching) in terms of accuracy or efficiency.
Magellan~\cite{2016-VLDB-Magellan.pdf} is the closest to us in spirit, as it also enables users to develop EM solutions end-to-end. However, like most existing solutions, Magellan is a supervised system that requires labeled data for training, while \systemx is weakly supervised that works on noisy LFs. In addition, in Magellan, users are in charge of many parts of the pipeline (e.g. labeling tuple pairs and training matchers), while in \systemx users only need to focus on developing LFs. However, because developers need to closely inspect EM data to develop LFs and iterate, it calls for an IDE that no existing EM solutions have looked at. 

Snorkel is a general-purpose data programming framework that is not tailored to EM, and does not have an IDE. In comparison, our \systemx system makes many optimizations to data-programming in the context of EM, such as pre-built library functions for EM, automatically-generated LFs, and optimized aggregation in the labeling model tailored to EM tasks.


\vspace{-1mm}
\section{Solution Overview}
We discuss the IDE in \systemx. We build \systemx's interface with Vue.js$^1$ and embed a Jupyter lab environment$^2$ for various coding support. We use flask$^3$ as our backend. ($^1$vuejs.org; $^2$jupyterlab. readthedocs.io; $^3$flask.palletsprojects.com)



\vspace{-2mm}
\subsection{Write, Debug, and Combine LFs}
\vspace{-1mm}
\stitle{1. Writing LFs}. Similar to a general-purpose programming IDE that provides various features (e.g. code completion and debugging) to assist users write programs, \systemx's IDE also provides essential support for developers to write LFs for EM. 

\textit{1.1 Smart data sampling.} To write LFs for EM, users need to examine tuple pairs from a specific EM task, in order to develop intuitions/heuristics that can be turned into code (LFs) to quickly label matches/non-matches. Randomly sampled pairs are likely non-matches in EM due to the class imbalance problem, and hence are not very useful. \systemx includes a smart data sampling strategy to show tuple pairs that are likely matches, but are not labeled as matches by the current labeling model. We use pre-trained semantic sentence models (e.g. sentence-BERT~\cite{sbert}) to obtain an embedding for each tuple, from which we use standard LSH to perform blocking. From pairs that remain after blocking, we compute their similarity scores and sample likely matches that are not currently as so.



\textit{1.2 Builtin utility functions for EM.} An LF for EM typically invokes standard utility functions such as: (1) text pre-processing (e.g., lower casing, stemming), (2) tokenization (e.g., 3-gram or space separation), (3) token weighting (e.g., equal weight or TF-IDF), and (4) distance functions (e.g., Jaccard distance or edit distance).  \systemx includes a library of utility functions along these four dimensions (described in detail in Figure 2 of our prior work~\cite{li2021autofuzzyjoin}), to make writing LFs easy.
We plan to expand the utility functions by including pre-trained matchers for specific entity types (e.g., People, Organization, Address, etc)~\cite{Autoem}, so that users can directly invoke pre-trained matchers relevant to their EM task in their LFs.


\textit{1.3 Automatically generated LFs.} Despite the aforementioned features in \systemx, for first-time users it may still be hard to write LFs, and even for experienced users it is still laborious to write many LFs. \systemx leverages our prior work \textsc{Auto-FuzzJoin}~\cite{li2021autofuzzyjoin} to automatically generate high-quality LFs tailored to given EM tasks, so that users can incorporate them directly without writing a single line of code.  Specifically, leveraging the fact that one of the input tables is likely a reference table with no or few duplicates (common in data warehouse setting~\cite{Chaudhuri} and shown in  ~\cite{li2021autofuzzyjoin} to hold on over 90\% EM datasets from~\cite{magellandata}), \textsc{Auto-FuzzJoin} can estimate the precision/recall of LFs and automatically generate high-quality LFs,  instantiated using different preprocessing functions, tokenization, weights, distance functions, and threshold values.


\stitle{2. Debugging LFs.} After users write LFs, they need to test/debug LFs to ensure that (1) LFs can be applied on tuple pairs without creating exceptions; and (2) LFs produce high-quality labeling results. To this end, \systemx supports the following two forms of debugging. 



\textit{2.1 Syntactical debugging.} To help users program LFs, we integrate the jupyter lab environment with a debugger extension in \systemx. This supports LF debugging just as in a general-purpose IDE: users can set breakpoints, step into LFs, execute code line by line, and see the values of the variables. 

\textit{2.2 Semantic debugging.} LFs that users write may produce incorrect labels on parts of the input data, and to help users improve LF's quality, \systemx has a labeling model to estimate each LF's false positive (FP) and false negative (FN) rate. Furthermore, in the IDE we provide an intuitive GUI where users can point and click to quickly narrow down to the record pairs where each LF may be making mistake. After examining these pairs, the user can iteratively improve the existing LF, or choose to write a new LF that complements with existing LFs. Each modification of LF will trigger FP and FN rates to be re-calculated, so that users can have a holistic view of the quality of all LFs.
\vspace{-1mm}
\stitle{3. Combining LFs.} Given a number of LFs (automatically generated or manually writte), users can trigger the labeling model in \systemx to combine signals from all LFs and produce an overall label. While existing labeling models (e.g., Snorkel~\cite{ratner2016data}) can be used here, \systemx develops an optimized EM-specific labeling model that leverages two properties unique to EM.

First, since EM problems typically have a class imbalance problem (the number of non-matches greatly exceeding the number of matches), using a single accuracy parameter to model LF quality (e.g.,~\cite{ratner2016data}) is insufficient. We develop two class dependent accuracy parameters, i.e. labeling accuracy $\alpha_M$ for matches and labeling accuracy $\alpha_U$ for non-matches. The accuracy parameters and the latent ground-truth label $y$ can then be estimated by an EM algorithm.

Second, tuple pairs in EM are not entirely independent due to the transitivity property, namely,  i.e. tuple pairs $(t_1, t_2)$ and $(t_1, t_3)$ both being match leads to $(t_2, t_3)$ being a match. To incorporate this transitivity property, we adopt the method in our prior work ZeroER~\cite{zeroer}. Specifically, we capture the transitivity property among any three tuples $t_i$, $t_j$ and $t_k$ by an inequality $\gamma_{i,j,M} \times \gamma_{i,k,M} \leq \gamma_{j,k,M}$ where $\gamma_{i,j,M}$ denotes the the probability of tuple pair $(t_i, t_j)$ being a match. All possible triples $t_i, t_j, t_k$ form a feasible set $Q$ for the probabilistic labels of the tuple pairs. 
We then enforce the transitivity constraint by projecting the estimated probabilistic labels to the feasible set $Q$ at each E-step. Our preliminary experiments on real-world benchmark datasets~\cite{erhardws} shows that our labeling model improves the F1-score of the state-of-the-art labeling model~\cite{ratner2017snorkel} by $12\%$ on average. We defer details of these results to a full paper on \systemx in the future.

\vspace{-5mm}
\subsection{Integrated and Intuitive GUI}
Similar to a general-purpose IDE, we provide an integrated development environment with an intuitive GUI for developing LFs, that allows users to visually inspect tuple pairs that are relevant to the given LFs. 
\systemx's GUI  has the following four main panels as shown in Figure~\ref{fig:panda_interface}(1).


\vspace{-1mm}
\stitle{EM Stats Panel.} This panel monitors the EM task's core statistics including table sizes, candidate set size, number of matches found, and estimated precision of the current solution. Users can click on the estimated precision to trigger the labeling process, where a sample of matches found will be loaded in the \textit{Data Viewer Panel} for users to label.

\vspace{-1mm}
\stitle{LF Stats Panel.} This panel monitors the core statistics of all LFs provided to \systemx so far. For every LF, it displays the name, number of matches/non-matches/abstains, and the estimated false-positive/ false-negative rates.
Users can also click on the header of any column to sort all LFs by values in the column.     
Users can click on the name of a LF to display its code snippet. In addition, users can also click on any actual statistics of any chosen LF to display the relevant tuple pairs in the \textit{Data Viewer Panel}. For example, in Figure~\ref{fig:panda_interface}(1), if the user clicks on 0.1402, the estimated false positive rate of the LF  name\_overlap, the \textit{Data Viewer Panel} will show all candidate pairs that the LF labels as +1, but the labeling model labels as -1. 

\vspace{-1mm}    
\stitle{Data Viewer Panel.} This panel displays the actual tuple pairs in a tabular format, where every row corresponds to a tuple pair. The first column titled "M/U" shows ground-truth labels (not available initially), and users may left/right click on the cells to provide match/non-match labels. 
The next set of columns show the actual data values for corresponding attributes in the two input tables. The data is presented in two sub-rows where the first row (colored in blue) shows data from the left table and the second row (colored in purple) shows data from the right table. This side-by-side comparison allows users to easily identify similarity/difference between the two tuples. 
When the user click on the "show" button, a sample of likely matches with label -1 or 0 from the current labeling model will be shown, each associated with a "likelihood" of matching score (in the last column) provided by the smart sampling method mentioned above. The user can sort the table by this column to quickly identify matches as show in Figure~\ref{fig:panda_interface}(1). 
    
\vspace{-1mm}
\stitle{LF Authoring Panel.} This panel embeds a JupyterLab environment for users to write and debug LFs.  After users click on the "load data" button, a notebook will be automatically generated with required dependencies imported in the first cell and automatically discovered LFs listed in the second cell right away, e.g. the auto\_lf\_0 function in Figure~\ref{fig:panda_interface}(1). Users can directly add new LFs or modify the existing auto LFs to create new LFs.
The debugger extension of jupyter lab is enabled to debug python LFs (breakpoints, step-in, etc.). The last cell of the notebook contains a single line of code that when triggered, combines existing LFs and updates the \textit{EM Stats Panel} and \textit{LF Stats Panel}. LFs are applied incrementally, i.e. only the new and modified LFs are executed. 

\vspace{-5mm}
\section{Demonstration Scenario}
\revise{\systemx operates in two phases: a \textit{development} phase and a \textit{deployment} phase. In the development phase, users focus on designing a set of LFs interactively. In the deployment phase, the final LFs are used to label the entire dataset.}
The LF development workflow of \systemx is shown in Figure~\ref{fig:panda_interface}(2). We demonstrate these steps using a real-world dataset Abt-Buy as follows:
\begin{figure*}[]
	\vspace{-10mm}
	\centering
	\includegraphics[clip, trim=1cm 5.7cm 4.5cm 1.2cm,width=17.8cm]{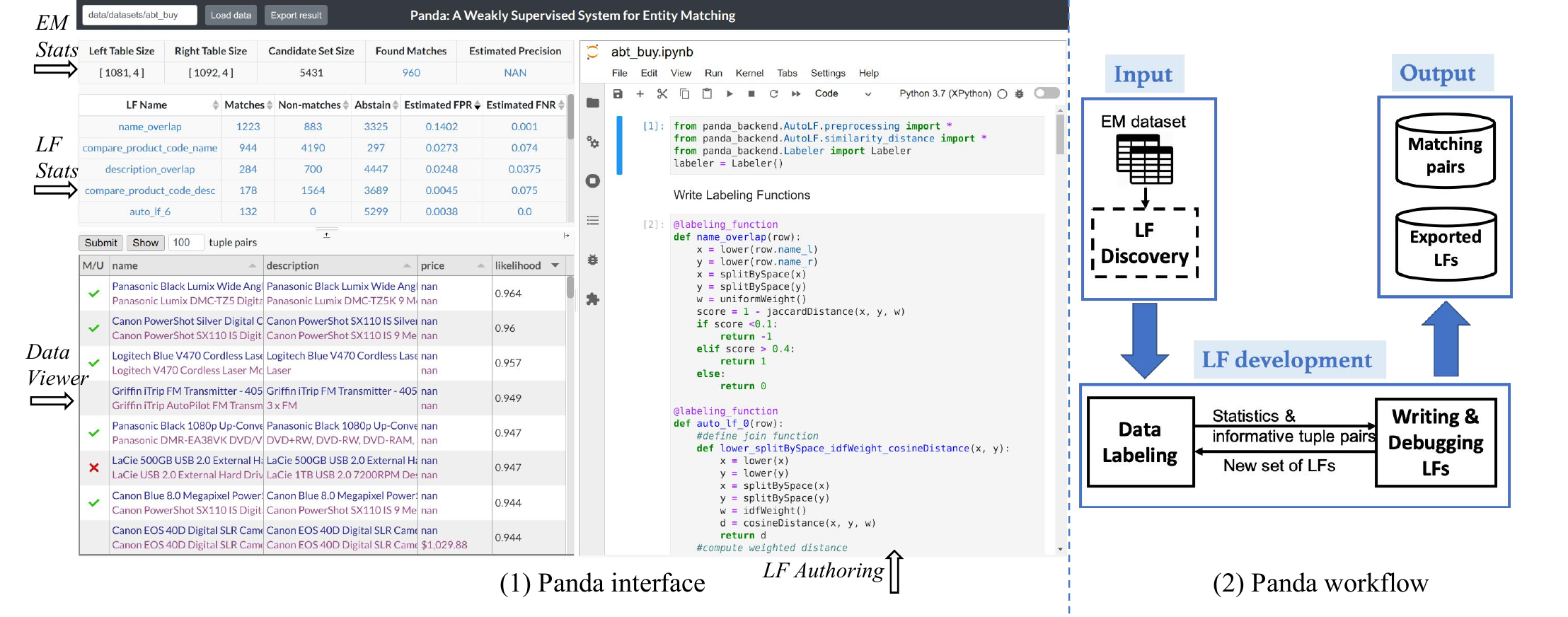}
	\vspace{-10mm}
	\caption{(1)User interface of Panda. (2)The LF Development Workflow in Panda. Dashed line denotes operations automatically done by the system; Red line denotes flow of debugging LF quality; Blue line denotes flow of discovering more matches; Purple text denotes data being sent to the next step.}
	\label{fig:panda_interface} 
	\vspace{-5mm}
\end{figure*}

\vspace{-1mm}
\stitle{Step 1: Uploading dataset and initialization.} The user uploads the dataset Abt-Buy by providing a file path and clicking on the "Load data" button. Then, the system performs blocking and discovers LFs automatically. A notebook that contains the discovered LFs (e.g. auto\_lf\_0 in Figure~\ref{fig:panda_interface}(1)) is generated, and the user opens it from the \textit{LF Authoring Panel}. The discovered LFs are combined by the labeling model to obtain EM \& LF stats (e.g. number of found matches), which are shown in the \textit{EM Stats Panel} and \textit{LF Stats Panel}.

\vspace{-1mm}
\stitle{Step 2: Viewing tuple pairs and coming up with LF ideas.} The user checks the current EM \& LF stats and wants to find more matches by manually writing more LFs.  The user clicks on the "Show" button. The system performs smart sampling and shows in \textit{Data Viewer Panel} some likely matching pairs that are abstained or labeled as non-match by the current LFs. For example, the \textit{Data Viewer Panel} in Figure~\ref{fig:panda_interface}(1) shows such pairs. The user sees that for tuple pairs that are matches, the "name" attribute likely have a high overlap of words. She can write an LF, name\_overlap, based on this idea.

\vspace{-1mm}
\stitle{Step 3: Writing LF.} The user goes to the \textit{LF Authoring Panel} to implement the idea. For preprocessing, she wants to convert everything to lower case and split input by spaces. She can navigate to the auto-generated LFs pre-populated in the same Jupyter notebook, and directly copy/paste examples of these built-in utility functions like "lower()" and "splitBySpace()" from auto\_lf\_0. She can further copy/paste examples of distance functions like "jaccardDistance()", from a different LF auto\_lf\_1.

Having finished writing LFs, the user executes the last cell in the notebook: "labeler.apply()". This incrementally applies the newly written LFs and updates the \textit{EM Stats Panel} and \textit{LF Stats Panel} accordingly. If the LFs have bugs causing errors, the user can turn on the debug mode of Jupyter Lab by clicking on the top-right toggle button. In the debug mode, user can debug the LF like in a normal Python IDE, e.g. creating breakpoints, etc. 

\vspace{-1mm}
\stitle{Step 4: Debugging LF quality.}
After a few iterations between Step 2 and Step 3, the user has written quite a few LFs and run out of ideas for new LFs. The user can now turn to inspect the overall precision and coverage of existing LFs. 

For example, in Figure~\ref{fig:panda_interface}(1), the user first clicks on "Estimated FPR" to sort the \textit{LF Stats Panel} by the false positive rate and finds that the FPR of LF name\_overlap is very high at $0.1402$. Next, the user can click on the value $0.1402$: the tuple pairs that this LF labels as 1 but the label model labels as -1 will be shown in \textit{Data Viewer Panel}. By examining the pairs, the user finds out that these false positive pairs do not have enough word overlapping. As a result, the user goes to the notebook and changes the threshold of being a match in LF name\_overlap from $>0.4$ to $>0.6$. After re-applying the LF, the FPR of the LF decreases to $0.0094$ and the user is satisfied.

\vspace{-1mm}
\stitle{Step 5: Estimate overall EM quality.} 
After a few iterations in Step 2, Step 3 and Step 4, the user has many high quality LFs. She is satisfied with the stats and wants to know the overall precision and recall. The user clicks on the value in the "Estimated Precision" column in the EM stats column (initialized as "NAN" in Figure~\ref{fig:panda_interface}(1)). The \textit{Data Viewer Panel} will show a random sample of matches predicted by the label model. The user can label these samples with left/right clicks in the first column to mark them matches/non-matches, after which an estimated precision will appear based on these human labels. As for recall, the user can inspect tuple pairs that are likely matches by clicking on the ``Show'' button, create LFs for those tuple pairs (using Step 2 and Step 3), until no more matches from the sample. This ensures that very few true-matches are missing and the overall recall is high.

\vspace{-2mm}
\section{Discussion and Future work}
We note that \systemx currently focuses on enabling developers to interactively author LFs and produce high-quality EM solutions. We plan to add features in \systemx so that it can handle large tables with millions of records, e.g., by down-sampling input data for LF development, which can then be applied to the entire dataset in a scale-out manner using infrastructures like Spark.
In current work, blocking is done by first obtaining a embedding vector for each tuple using a pretrained sentence model~\cite{sbert} and then using LSH of the embedding vectors to obtain candidate set of tuple pairs.

\vspace{-2mm}
\bibliographystyle{ACM-Reference-Format}
\bibliography{sample}

\end{document}